\documentclass[twocolumn,english,prl,aps,superscriptaddress,amsmath,amssymb,floatfix]{revtex4-2}
\usepackage[T1]{fontenc}
\usepackage{verbatim}
\usepackage{amsmath}
\usepackage{amssymb}
\usepackage{graphicx}

\makeatletter
\usepackage{times}
\usepackage{textcomp}
\usepackage{epstopdf}
\usepackage{braket}
\usepackage{tikz}
\usepackage[colorlinks,linkcolor=blue,citecolor=blue]{hyperref}
\usepackage{tikz-network}
\usepackage{amsfonts}
\newtheorem{theorem}{Theorem}

\pdfpageheight\paperheight
\pdfpagewidth\paperwidth

\@ifundefined{textcolor}{}{%
 \definecolor{BLACK}{gray}{0}
 \definecolor{WHITE}{gray}{1}
 \definecolor{RED}{rgb}{1,0,0}
 \definecolor{GREEN}{rgb}{0,1,0}
 \definecolor{BLUE}{rgb}{0,0,1}
 \definecolor{CYAN}{cmyk}{1,0,0,0}
 \definecolor{MAGENTA}{cmyk}{0,1,0,0}
 \definecolor{YELLOW}{cmyk}{0,0,1,0}
}

\usepackage{xcolor}\usepackage{soul}
\def \Tr {\mathrm{Tr}}
\definecolor{blue}{rgb}{0,0,1}
\definecolor{red}{rgb}{1,0,0}
\definecolor{green}{rgb}{0,1,0}

\usepackage{babel}
\begin{document}
\title{Free-fermion 
Page Curve: Canonical Typicality and Dynamical Emergence}
\author{Xie-Hang Yu}
\affiliation{Max-Planck-Institut f\"ur Quantenoptik, Hans-Kopfermann-Stra{\ss}e 1, D-85748 Garching, Germany}
\affiliation{Munich Center for Quantum Science and Technology, Schellingstra{\ss}e 4, 80799 M\"unchen, Germany}
\author{Zongping Gong}
\affiliation{Max-Planck-Institut f\"ur Quantenoptik, Hans-Kopfermann-Stra{\ss}e 1, D-85748 Garching, Germany}
\affiliation{Munich Center for Quantum Science and Technology, Schellingstra{\ss}e 4, 80799 M\"unchen, Germany}
\author{J. Ignacio Cirac}
\affiliation{Max-Planck-Institut f\"ur Quantenoptik, Hans-Kopfermann-Stra{\ss}e 1, D-85748 Garching, Germany}
\affiliation{Munich Center for Quantum Science and Technology, Schellingstra{\ss}e 4, 80799 M\"unchen, Germany}
\begin{abstract}
We provide further analytical insights into the newly established noninteracting (free-fermion) Page curve, focusing on both the kinematic and dynamical aspects. First, we unveil the underlying canonical typicality and atypicality for random free-fermion states
. The former appears for a small subsystem and is exponentially weaker than the well-known result in the interacting case. The latter explains why the free-fermion Page curve differs remarkably from the interacting one when the subsystem is macroscopically large, i.e., comparable with the entire system. Second, we find that the free-fermion Page curve emerges with unexpectedly high accuracy in some simple tight binding 
models in long-time quench dynamics. This contributes a rare analytical result concerning quantum thermalization on a macroscopic scale, where conventional paradigms such as the generalized Gibbs ensemble and quasi-particle picture are not applicable. 
\end{abstract}
\maketitle

\emph{Introduction.--}As
a central concept in quantum information science \citep{nielsen00},
entanglement has been recognized to play vital roles in describing
and understanding quantum many-body systems in and out of equilibrium
\citep{Luigi2008,relation_entropy_Phase,Eisert2015,Abanin2019}. 
For example, entanglement area laws for ground states
of gapped local Hamiltonians enable their efficient descriptions based
on tensor networks \citep{RevModPhys.93.045003}, 
while their violations may signature 
quantum phase transitions \citep{Vidal2003,Calabrese2004}. 
The emergence of thermal ensemble from unitary evolution, a process
known as quantum thermalization \cite{Srednicki1994}, is ultimately attributed to the entanglement
generated between a subsystem and the complement \citep{Nandkishore2015}. 

Almost thirty years ago, Page considered the fundamental
problem of bipartite entanglement in a fully random many-body system
and found a maximal entanglement entropy (EE) up to finite-size corrections \cite{Page1993}.
This seminal work was originally motivated by the black-hole information
problem \cite{Page1993blackhole}. 
Remarkably, it
has been attracted increasing and much broader interest in the past
decade, 
due not only 
to the new theoretical insights from quantum thermalization \citep{PhysRevLett.115.267206,PhysRevLett.119.220603,Nakagawa2018,PageCurve_Thermal,Lu2019,PageCurve_Thermal2,PhysRevLett.125.021601,Kaneko2020,PhysRevB.91.081110} and 
quantum chaos \citep{Sekino_2008,
Chaos_Scrambl,Nahum2018},
but also to the practical relevance in light of the rapid experimental development in quantum simulations 
\citep{entropy_measure1,entropy_measure2,entropy_measure3,long_range_accessible,long_range_accessible2,Semeghini2021,
Yang2020,
Liu2022}. 
In particular, the saturation of maximal entropy has been found to
be a consequence of canonical typicality \citep{Popescu2006,Goldstein2006,Reimann2007},
which means most random states behave locally like the canonical ensemble. 
This typicality 
behavior
has been argued to emerge in generic interacting many-body
systems satisfying the eigenstate thermalization hypothesis \citep{Srednicki1994,QuantumChaosETH,Abanin2019,
Rigol2008,Nandkishore2015,Moessner2017} 
and can even be rigorously established or ruled out in specific situations \cite{
ETH_CT_proof3,Hamazaki2018}. 

\begin{figure}
\includegraphics[width=1\columnwidth]{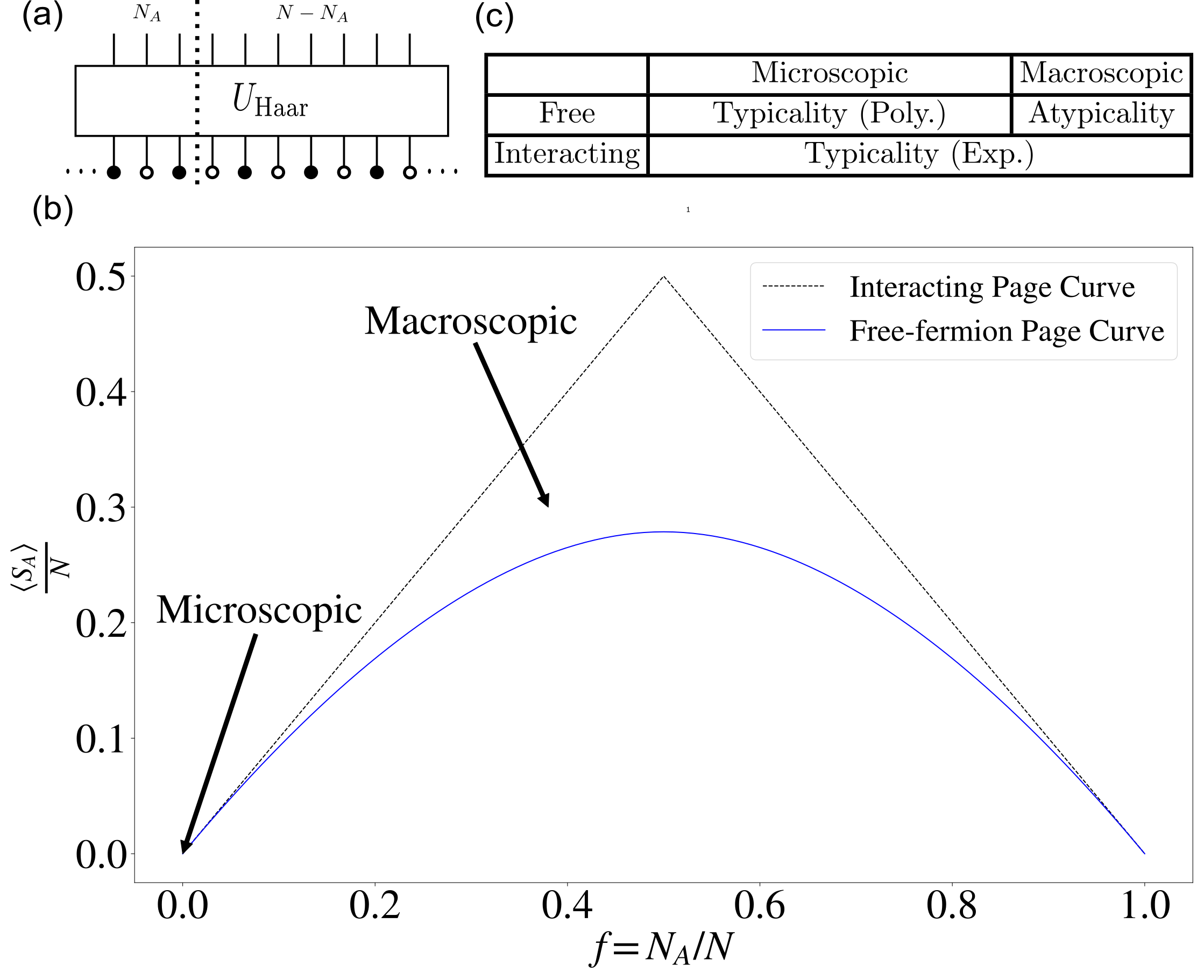}\caption{(a) The entire 
free-fermion system has $N$ sites with half filling. The 
subsystem of interest 
has $N_{A}$ ($N_{A}\le
N$) sites. 
The RFG ensemble is generated by Haar-random Gaussian unitaries with number conservation. 
(b) 
The Page curves of the RFG 
and interacting ensemble in the thermodynamic limit $N\to\infty$. 
It is obvious that these two Page curves agree with each other in the microscopic region but show a $\mathcal{O}(1)$ deviation in the macroscopic region. The interacting Page curve in the thermodynamic limit is always saturated. (c) The table summarizes the typicality/atypicality property for the RFG and interacting ensembles. Here ``Poly." and ``Exp." indicates polynomial and exponential scalings, respectively.}
\label{Setup_of_CCRFG_ensemble}
\end{figure}

In this Letter, we provide analogous insights into
the noninteracting counterpart of Page's problem.  
That is, we focus on free fermions or (fermionic) Gaussian states, which are of their own interest in quantum many-body physics, quantum information and computation \citep{matchgates,PhysRevA.65.032325,Bravyi2005,Wolf2006,Banuls2007,Fidkowski2010,PhysRevLett.116.030401,Shi2018,PhysRevLett.120.190501,PhysRevLett.121.200501,Circuit_complexity_free_fermion,fermionicTomograph1,Oszmaniec2022,Matos2022,PhysRevLett.119.020601,PhysRevB.100.165135,PhysRevB.106.035143}. Somehow surprisingly, in the seemingly simpler noninteracting
case, the subsystem-size dependence of averaged EE,
which is described by the Page curve pictorially, was not solved until
very recently \citep{Bianchi2021,Bianchi2021a,PhysRevB.104.214306}. It turns out to be similar to the interacting case for a small subsystem, but differ significantly otherwise. See Fig.~\ref{Setup_of_CCRFG_ensemble}(b) for an illustration. With the measure concentration results on compact-group manifolds,  we establish the corresponding 
canonical typicality (atypicality) in microscopic (macroscopic) regions for the free-fermion ensemble. Thus we explicitly explain the similarity and difference from the kinematic aspect. In addition, we show that the free-fermion Page curve can be relevant to extremely simple tight-binding models via long-time quench dynamics. By classifying the systems according to their conserved (eigen) mode occupation numbers, we construct two classes of Hamiltonians which can/cannot give rise to a highly similar Page curve. Our finding concerning macroscopic properties which cannot be captured by the generalized Gibbs ensemble or quasi-particle picture and thus goes beyond the conventional paradigm of local thermalization.

\emph{Canonical typicality and atypicality.--}
We start by generalizing 
the main result in \citep{Popescu2006} to the random fermionic Gaussian (RFG) ensemble. While \citep{Popescu2006} already considers possible restrictions, we stress that Gaussianity is inadequate since Gaussian states do not constitute a Hilbert subspace. 
For simplicity, we 
consider number-conserving systems with totally $N$ modes occupied by $N/2$ fermions, i.e., the half-filling case. Compared to the fully random case without number conservation, this setting appears to be more physically comprehensible and experimentally relevant, while displaying exactly the same Page curve 
\citep{Bianchi2021a,Bianchi2021}. 
More general ensembles are discussed 
in Supplemental Material \cite{SM}. 

A pictorial illustration of our setup is shown 
in Fig.~\ref{Setup_of_CCRFG_ensemble}(a). 
Due to Wick's theorem \citep{Hackl2021}, a fermionic Gaussian 
state $\rho$ is 
fully captured 
by its covariance matrix $C_{j,j'}=\mathrm{Tr}(\rho a_{j}^{\dagger}a_{j'})$ \cite{Peschel2003}.
Here $a_{j}$ is the annihilation operator for mode $j$, which may label, e.g., a lattice site. 
As the covariance matrix for any RFG-pure state can be related to each
other by a unitary transformation, 
the uniform distribution over this ensemble can be generated at the
level of the covariance matrix $\{C=UC_{0}U^{\dagger}\}$. Here 
$U$ is taken Haar-randomly over the unitary group $\mathbb{U}(N)$ 
\citep{Bianchi2021,Bianchi2021a} and
$C_{0}$ is an arbitrary reference 
covariance matrix in the ensemble satisfying $C_0^2=C_0$ and $\Tr C_0=N/2$. 

An important property of Gaussian states is that their subsystems remain 
Gaussian. 
We denote $C_{A}$ as 
the $N_A\times N_A$ covariance matrix restricted
to 
subsystem $A$ with $N_A$ modes. The EE $S_A=-\Tr(\rho_A\log_2\rho_A)$ of the reduced state $\rho_A=\Tr_{\bar A}\rho$ ($\bar A$: complement of $A$) then reads:
\begin{equation}
\begin{split}
    S_{A}=&-\Tr(C_A\log_2 C_A)\\
    &-\Tr((I_A-C_A)\log_2 (I_A-C_A)), 
    \end{split}
    \label{eq:SA}
\end{equation}
where $I_A$ is the identity matrix with dimension $N_A$.

Our first result is the measure concentration property of the covariance
matrix for RFG ensemble:

\begin{theorem}
For arbitrary $\epsilon>0$ and subsystem $A$, the probability that the reduced covariance matrix of a state in the RFG ensemble deviates from the ensemble average satisfies 
\begin{equation}
\mathbb{P}(d_{\rm HS}(C_{A},
I_{A}/2)\geq\eta+2\epsilon)\leq2e^{-\frac{\epsilon^{2}}{\eta'}}
\label{eq:Canonicality_result_in_CCRFG}
\end{equation}
and
\begin{equation}
\mathbb{P}(d_{\rm HS}^2(C_{A},
I_{A}/2)\leq\eta_{\rm a}-2\epsilon)\le 2e^{-\frac{\epsilon^2}{\eta'_{\rm a}}}    
\label{eq:atypicality}
\end{equation}
with $\eta=\sqrt{\frac{N_{A}^{2}}{2(N-1)}}$, 
$\eta'=\frac{12}{N}$, $\eta_{\rm a}=\frac{N_A^2}{4(N+1)}$, $\eta'_{\rm a}=\frac{12N_A}{N}$ and $d_{\rm HS}(C,C')=\sqrt{\Tr(C-C')^2}$ being the Hilbert-Schmidt distance. 
\end{theorem}
The proof largely relies on the generalized Levy's lemma
for Riemann manifolds with positive curvature \citep{measure_concentration1,measure_concentration2,Meckes2019},
which allows us to turn the upper bound on 
the distance average $\langle d_{\mathrm{HS}}(C_{A},I_{A}/2)\rangle\leq\sqrt{\frac{N_A^2}{2(N-1)}}$ or the lower bound $\langle d^2_{\mathrm{HS}}(C_{A},I_{A}/2)\rangle\geq\frac{N_A^2}{4(N+1)}$
into a probability inequality \citep{SM}. 
From Eq. (\ref{eq:Canonicality_result_in_CCRFG}) we can easily see for
infinite environments $N\to\infty$, the local microscopic system will have maximal entropy
$S_{A}\to N_{A}$.

We emphasize that in Eq. (\ref{eq:Canonicality_result_in_CCRFG}),
$\eta$ and $\eta'$ only scales polynomially with the (sub)system size. 
This contrasts starkly with the exponential scaling canonical typicality 
for random interacting 
ensemble \citep{Popescu2006}. 
Intuitively, this is because in the interacting 
case, the 
Hilbert-space dimension scales exponentially with the (sub)system size, 
which, however, simply equals to the size 
of the covariance matrix in the free-fermion case. 
Physically, the Gaussian constraint 
makes the ensemble
only explore a very limited sub-manifold in the entire 
Hilbert space.
This polynomial scaling means that, for a fixed subsystem size $N_A$, the reduced state still exhibits canonical typicality, while the atypicality is only polynomially suppressed by the environment size. Accordingly, the averaged EE should achieve the maximal value but with a polynomial finite-size correction. In fact, such an exponentially weaker canonical typicality (\ref{eq:Canonicality_result_in_CCRFG}) can also be exploited to explain the qualitatively larger variance of the EE for the RFG ensemble, which is $\mathcal{O}(N^{-2})$ 
in comparison to  
$e^{-\mathcal{O}(N)}$ in the interacting case 
\citep{Bianchi2021a,Bianchi2021,SM}.  

On the other hand, 
if the subsystem is macroscopically large, meaning that $f=N_A/N$ is $\mathcal{O}(1)$, the concentration inequality (\ref{eq:Canonicality_result_in_CCRFG}) becomes meaningless since $\eta$ is $\mathcal{O}(\sqrt{N})$, the same order as the Hilbert-Schmidt norm of $C_A$. Instead, we may take $\epsilon=\mathcal{O}(N^\alpha)$ with $\alpha\in(0,1)$ in Eq.~(\ref{eq:atypicality}), finding that the majority of the reduced covariance matrix differs significantly from the ensemble average. In other words, the RFG ensemble exhibits canonical atypicality in this case. In particular, this result implies an $\mathcal{O}(1)$ deviation in the EE density from the maximal value. We recall that, in stark contrast, the canonical typicality for interacting states is exponentially stronger and persists even on any macroscopic scale with $f<1/2$. 


The above discussions can be made more straightforward 
by considering the measure concentration property of 
$S_{A}$. By bounding $S_A$ using $d_{\rm HS}(C_A,I_A/2)$ from both sides, we obtain \cite{SM} 
\begin{equation}
\mathbb{P}(S_{A}\leq N_{A}-\epsilon)\le2e^{-\frac{(\sqrt{\epsilon}-\xi)^{2}}{\xi'}},\;\;\;\;\forall\epsilon>\xi^2
\label{eq:main_text_typicality_for_entropy}
\end{equation}
in the microscopic region and
\begin{equation}
\mathbb{P}(S_{A}\geq N_{A}-\xi_{a}+\epsilon) \leq2e^{-\frac{\epsilon^{2}}{\xi'_{a}}},\;\;\;\;
\forall\epsilon>0
\label{eq:main_text_atypicality_for_entropy}
\end{equation}
in the macroscopic region. Here $\xi=\sqrt{\frac{2N_{A}^{2}}{N-1}}$, $\xi'=\frac{192}{N}$, $\xi_{a}=\frac{N_{A}^{2}}{2\ln2(N+1)}$ and $\xi'_{a}=\frac{192N_{A}}{\ln^22N}$. 
Note that Eq.~(\ref{eq:main_text_typicality_for_entropy}) also becomes meaningless in the macroscopic region since $\xi^2$ will be comparable with $N_A$.
Choosing $\epsilon=(\xi+\mathcal{O}(N^{-\alpha/2}))^2$ for Eq.~(\ref{eq:main_text_typicality_for_entropy}) and $\epsilon=\mathcal{O}(N^\alpha)$ in Eq.~(\ref{eq:main_text_atypicality_for_entropy}) with $\alpha\in(0,1)$, we fully explain the microscopic similarity and macroscopic difference 
between 
the Page curves for the 
RFG and interacting ensembles. 
See Fig.~\ref{Setup_of_CCRFG_ensemble}(b) and (c).

\begin{figure*}
\includegraphics[width=0.9\textwidth]{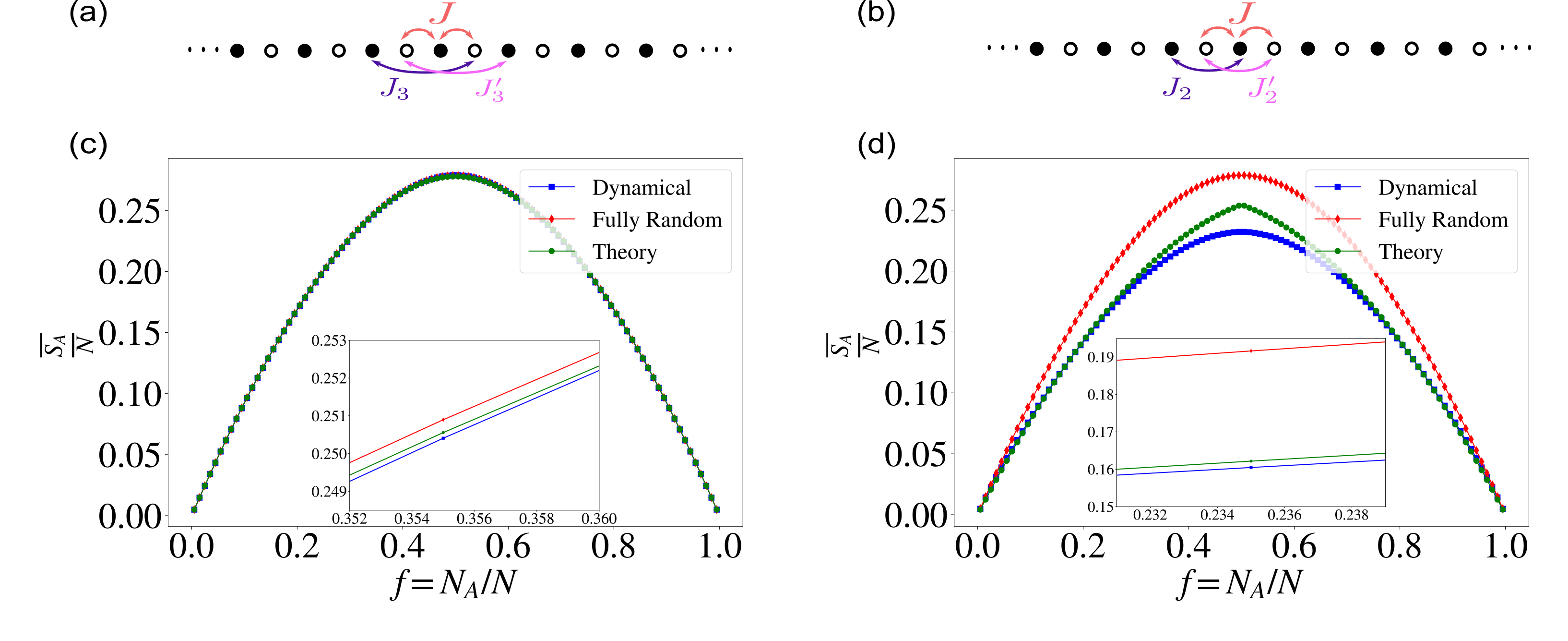}
\caption{(a) and (b) show the tight binding Hamiltonians $H_0+H_1$ with period 2. 
$H_1$ in (a) only includes the odd-range hopping, while (b) includes even-range hopping. (c) Dynamical Page curve for the minimal model (\ref{eq:NNH_Hamiltonian}) 
(blue) as a representative of (a) 
and its comparison with the Page curve for the RFG ensemble
(red) 
as well as our theoretic result up to order $\mathcal{O}(f^5)$ 
(green). 
Here $N=200$. 
These three lines are very close to each other, with a difference $\sim
10^{-3}$ which agrees with our analysis. 
This figure can also represent the general dynamical Page curve for Hamiltonians in (a). 
(d) Dynamical Page curve for Hamiltonian
$H=(\sum_{j=1}^{N}a_{j}^{\dagger}a_{j+1}+0.3\sum_{j:\mathrm{even}}a_{j}^{\dagger}a_{j+2}-0.3\sum_{j:\mathrm{odd}}a_{j}^{\dagger}a_{j+2})+{\rm H.c.}$ as a representative of (b). Here also $N=200$. The dynamical Page curve is obviously different
from the Page curve for the RFG ensemble. The considerable 
deviation between the
theoretical result and the dynamical Page curve near $f=\frac{1}{2}$, where higher-order terms become least negligible, is because we only calculate up to the third term in Eq. (\ref{eq:Taylor_expansion_for_entropy}) \cite{SM}. 
}
\label{dynamical_page_curve_total}
\end{figure*}

\emph{Dynamically emergent Page curve.--}
We recall that a particularly intriguing point of the (interacting) Page curve is its emergence in physical many-body systems with local interactions \cite{PhysRevLett.115.267206,PhysRevLett.119.220603,Nakagawa2018,PageCurve_Thermal,Lu2019,PageCurve_Thermal2}, which are typically chaotic but yet far from fully random. Indeed, a popular phenomenological theory for describing generic entanglement dynamics on the macroscopic level, the so-called entanglement membrane theory \cite{Nahum2018}, explicitly assumes that the entanglement profile of the thermalized system follows the Page curve. The intuition is that a long-time evolution can generate highly non-local correlations in a state and roughly exhaust the whole Hilbert (sub)space, provided the dynamics is ergodic. It is thus natural to ask whether the free-fermion Page curve could be relevant to thermalization in real physical systems without interactions. Note that this question is complementary to the aforementioned (a)typicality results, which are kinematic, i.e., irrelevant to dynamics, as in the interacting case \cite{Popescu2006}. 

We try to address the above question by 
analytically investigating the long-time averaged EE in the quench dynamics governed by 
some simple local quadratic Hamiltonians with number conservation. Hereafter, we use the term ``dynamical Page curve" to refer to this long-time averaged entanglement profile. 
Unlike \citep{PhysRevE.104.014146, Dias2021} which deal with models with strong spatiotemporal disorder so the emergence of the RFG Page curve is somehow expectable, we assume the Hamiltonian $H$ to be time-independent, translation-invariant (under the periodic boundary condition) and specify our initial state $|\Psi_0\rangle$ to be 
a period-2 density wave with half filling. Our simple setup thus appears to be far-from-random and highly experimentally accessible. See Fig.~\ref{dynamical_page_curve_total}(a-b) for a schematic illustration. The dynamical Page curve is 
formally given by $\overline{S(\rho_A(t))}$, where $\rho_A=\Tr_{\bar A}[e^{-iHt}|\Psi_0\rangle\langle\Psi_0|e^{iHt}]$ and $\overline{f(t)}=\lim_{T\to\infty}T^{-1}\int^T_0 dt f(t)$ denotes the long-time average. It is worth mentioning that the dynamical Page curve is ensured to be concave by translation invariance, as a result of the strong subadditivity of quantum entropy \cite{Wolf2008}. 

We primarily focus on the minimal 
model, i.e., a one-dimensional lattice with nearest-neighbor hopping:
\begin{equation}
H_{0}=\sum_{j}a_{j}^{\dagger}a_{j+1}+{\rm H.c.},
\label{eq:NNH_Hamiltonian}
\end{equation}
whose 
band dispersion reads $E_k=2\cos k$. We believe that the exact results for the large (spatiotemporal) scale dynamical behaviors of this fundamental model are interesting on their own.  Moreover, our method and results actually apply to much broader situations, as will soon become clear below. 

Surprisingly, despite the additional translation-invariant and energy-conserving constraints compared to the RFG ensemble, this minimal model (\ref{eq:NNH_Hamiltonian}) turns out to give rise to 
a dynamical Page curve extremely close to that for the 
RFG ensemble 
(see blue and red curves in Fig.~\ref{dynamical_page_curve_total}(c)). 
To gain some analytic insights, we perturbatively expand 
the entropy expression (\ref{eq:SA}) 
around $C_{A}=\frac{I_{A}}{2}$, obtaining 
\begin{equation}
S_{A}(t)=N_{A}-
\sum_{n=1}^{\infty}\frac{\mathrm{Tr}(2C_{A}(t)-I_{A}){}^{2n}}{2n(2n-1)\ln2}.
\label{eq:Taylor_expansion_for_entropy}
\end{equation}
Thanks to the translational invariance, $C_{A}(t)$ can be related to the block-diagonal momentum-space covariance matrix $\tilde C(t)=\bigoplus_k\tilde C_k(t)$ via 
$C_{A}(t)=\Pi_{A}U_{\rm F}\tilde{C}(t)U^{\dagger}_{\rm F}\Pi_{A}^{\dagger}$. Here
$U_{\rm F}$ and $\Pi_{A}$ are the Fourier transformation matrix and projector 
to subsystem $A$, respectively. 
The off-diagonal
elements of a $2\times2$ block $\tilde{C}_k(t)$ involve 
a time-dependent 
phase
$e^{i\theta_{k}(t)}$ with 
$\theta_{k}(t)=t(E_{k}-E_{k+\pi})$.
When calculating $\overline{\mathrm{Tr}(2C_{A}(t)-I_{A})^{2n}}$, 
we will encounter 
terms like $\overline{e^{i\theta_{k}(t)}e^{i\theta_{k'}(t)}}$,
which equals to $\delta_{k,k'+\pi}$ in the thermodynamic limit.
This contraction allows us to establish a set of Feynman rules for 
systematically calculating Eq.~(\ref{eq:Taylor_expansion_for_entropy})
order by order \cite{SM}. 

Since the bipartite EE is identical for either of the subsystems, the Page curve is reflection-symmetric with respect to $f=\frac{1}{2}$ and thus 
 it suffices to 
focus on $f=N_A/N\leq\frac{1}{2}$. In the thermodynamic limit, the dynamical Page curve turns out to be \cite{SM} 
\begin{equation}
\frac{\overline{S_{A}}}{N}=f-\frac{1}{\ln{2}}\left(\frac{1}{2}f^{2}+\frac{1}{6}f^{3}+\frac{1}{10}f^{4}\right)+\mathcal{O}(f^{5}).\label{eq:the_Page_curve_for_NNH}
\end{equation}
 On the other hand, the 
 Page curve for RFG ensemble
is \citep{Bianchi2021}
\begin{equation}
\frac{\langle S_{A}\rangle}{N}=f-\frac{1}{\ln{2}}\left(\frac{1}{2}f^{2}+\frac{1}{6}f^{3}+\frac{1}{12}f^{4}\right)+\mathcal{O}(f^{5}).\label{eq:the_Page_curve_result_for_CCRFG}
\end{equation}
The above two equations differ 
only by $\frac{1}{60\ln{2}}f^{4}+\mathcal{O}(f^5)$, 
which is as small as about $
10^{-3}$ even for $f$ near $1/2$.

Interestingly, if we add a perturbation $H_{1}$ to Eq.~(\ref{eq:NNH_Hamiltonian}),
as long as $H_{1}$ is period-2 
and 
only includes 
odd-range hopping, as represented by Fig.~\ref{dynamical_page_curve_total}(a),  
the dynamical Page curve can be analytically demonstrated 
to be the same as Eq. (\ref{eq:the_Page_curve_for_NNH}) in the thermodynamic 
limit, as the same Feynman rules apply \cite{SM}. 
One example 
is $H_{1}=J(\sum_{j:\mathrm{even}}a_{j}^{\dagger}a_{j+2m+1}-\sum_{j:\mathrm{odd}}a_{j}^{\dagger}a_{j+2m+1})+{\rm H.c.}$
for arbitrary $J$ and integer $m$. Thus, we have defined another ensemble
of fermionic Gaussian states by dynamical evolution, which covers a
wide class of Hamiltonians and this ensemble has remarkably similar
Page curve as the RFG ensemble. 

However, if $H_{1}$ includes even-range 
hopping, as represented by Fig.~\ref{dynamical_page_curve_total}(b) 
the dynamical Page curve will be very different, 
as shown in Fig.~\ref{dynamical_page_curve_total}(d).
This can be easily explained with the canonical typicality property
proved above: for this class of Hamiltonians, their conserved (eigen) mode
occupation number $n_{k}$ deviates from the average value of RFG ensemble,
which is 
$\frac{1}{2}$. 
Thus, 
the dynamical ensemble is naturally 
``atypical" even for microscopic scale because the local conserved observable is constructed from mode occupation numbers \citep{Ishii2019}. This result implies the reduced state on a small subsystem deviates considerably from being maximally mixed so that the tangent slope of the dynamical Page curve at $f=0$ is well below that for the RFG ensemble.
In contrast, one can show that all the conserved mode occupation number for the class of Hamiltonians mentioned in the last paragraph are $\frac{1}{2}$. 

All the observations above constitute our second main result: 
\begin{theorem}
The RFG ensemble-like dynamical Page curve (\ref{eq:the_Page_curve_for_NNH}) emerges for a period-2 short-range free-fermion Hamiltonian if and only if the conserved mode occupation numbers are $1/2$. 
\end{theorem}



\emph{Discussions.--}
It is well-known that the generalized Gibbs ensemble (GGE) 
characterizes 
the local thermalization of integrable systems including free fermions \citep{PhysRevLett.98.050405, Cassidy2011,doi:10.1126/science.1257026, Essler2016,Ishii2019}. 
However, in principle,
GGE only predicts the expectation values of observables, which do not include the entropy. Note that the former (latter) is linear (nonlinear) in $\rho_A$ and thus commmutes (does not commute) with time average. 
Moreover, 
we also study the macroscopic scale, which can not be captured by GGE as well as its recently proposed refined version \cite{Lucas2022} concerning the purified subsystem by measuring the complement 
\cite{Ho2022}. In this sense, our study goes well beyond the conventional paradigm of quantum thermalization in integrable systems, pointing out especially the highly nontrivial behaviors on the macroscopic level, where typicality may completely break down. 

Finally, let us mention the relation between our 
strategy
and the quasi-particle picture, which is widely 
used to calculate 
EE 
growth \citep{PhysRevLett.127.060404,Jurcevic2014,Castro2016_,Essler2016,Calabrese2005,Fagotti2008,Bertini2018,BertiniB2018_2}. 
It turns out this picture fails to 
reproduce the 
dynamical Page curve. Under the 
periodical boundary condition, the quasi-particle
picture predicts 
$\overline{S_{A}}=N-\frac{N_{A}^{2}}{N}$
for the Hamiltonian satisfying the conditions in Theorem 2 \cite{SM}. This result is obtained by counting the steady number of entangled pairs shared by $A$ and $\bar A$. 
On the other hand, noting that $(2C_{A}-I_{A})^{2n}\leq(2C_{A}-I_{A})^{2}$,
if we replace all the higher-order terms of $(2C_{A}-I_{A})$ in Eq. (\ref{eq:Taylor_expansion_for_entropy})
with $(2C_{A}-I_{A})^{2}$, we will get a lower entropy 
bound, 
which coincides with the prediction 
by the quasi-particle picture: $S_{A}\geq N_{A}-\frac{\mathrm{Tr}(2C_{A}-I_{A})^{2}}{\ln2}\sum_{n}\frac{1}{2n(2n-1)}=N_{A}-\frac{N_{A}^{2}}{N}$.
It is thus plausible 
to argue that the quasi-particle picture ignores
possible higher-order correlations beyond 
quasi-particle pairs. 

\emph{Conclusion and outlook.--}
We have derived the canonical
(a)typicality for the RFG ensemble 
and pointed out the quantitative 
scaling difference in atypicality suppression 
from 
interacting systems. This 
explains the very different behaviors of the Page curves. 
We have also explored the relevance to long-time quench dynamics of 
free-fermion systems. 
To our surprise, some simple time-independent
Hamiltonians are enough to 
make 
the free-fermion Page curve 
emerge to a very high accuracy. 
We 
analytically prove a necessary and sufficient condition about this behavior. The breakdown 
of the quasi-particle picture was also discussed.

Strictly speaking, we define a new 
ensemble arising from 
a wide class of free-fermion Hamiltonians, whose dynamical Page curve resembles a lot but yet differs from 
the fully random one. 
The properties of this new ensemble and its corresponding Page curve 
merit further study. Another interesting question is how the 
dynamical Page curves will be enriched upon imposing 
additional symmetries (such as the Altland-Zirnbauer symmetries \cite{Altland1997}), in which case one may naturally consider 
the symmetry-resolved EE 
\citep{PhysRevD.106.046015,Lau2022}. 
Our work proposes a methodology to study this question. Besides, whether
or not the fully random Page curve can emerge exactly for a time-independent free Hamiltonian 
also remains open.

We thank L. Piroli for valuable communications. Z.G. is supported by the Max-Planck-Harvard Research Center for Quantum Optics (MPHQ). J.I.C. acknowledges support by the EU Horizon 2020 program through the ERC Advanced Grant QENOCOBA No. 742102.

\emph{Note added.---}While finalizing this manuscript, a related
work by Isoue \emph{et al}. appeared in Ref.~\cite{Iosue2022}, which reported the 
typicality for random bosonic Gaussian states.

\bibliographystyle{utphys}

\bibliographystyle{apsrev4-2}
\addcontentsline{toc}{section}{\refname}
\bibliography{MyCollection}
\clearpage{}

\onecolumngrid
\renewcommand{\thefigure}{S\arabic{figure}}
\setcounter{figure}{0} 
\renewcommand{\thepage}{S\arabic{page}}
\setcounter{page}{1} 
\renewcommand{\theequation}{S.\arabic{equation}}
\setcounter{equation}{0} 
\setcounter{section}{0}

\begin{center}
\textbf{\textsc{\LARGE{}Supplementary Information}}{\LARGE\par}
\par\end{center}

\tableofcontents{}

In this Supplemental Mateiral, we provide detailed proof of Theorem 1 and 2 in the main text. We also provide the calculations of other results and conclusions in the main text and discuss their generalization.

\section{Proof of Canonical Typicality/Atypicality for the RFG ensemble}

In this section, we consider number conserving fermionic
Gaussian ensemble with $N$ modes occupied by $m$ fermions. The
notation follows the main text. In particular, 
$\langle\cdots\rangle$ is
used to denote the average value over the ensemble. The covariance
matrix of the subsystem $A$ for a particular random Gaussian state
is 
\begin{equation}
C_{A}=\Pi_{A}UC_{0}U^{\dagger}\Pi_{A}^{\dagger},
\end{equation}
where $\Pi_{A}$ is the projection operator on the the subsystem with size
$N_{A}\times N$, $U$ 
is taken Haar randomly over $\mathbb{U}(N)$ and
$C_{0}$ satisfies $C_{0}^{2}=C_{0}$ and $\Tr C_{0}=m$. In the following,
the distance between two matrices is measured by Hilbert-Schmidt distance
$d_{\mathrm{HS}}$. We define a function $f:\mathbb{U}(N)\to\mathbb{R}$
as
\begin{equation}
f(U)=d_{\mathrm{HS}}(\Pi_{A}UC_{0}U^{\dagger}\Pi_{A}^{\dagger},\langle C_{A}\rangle).
\end{equation}
It is easy to check that $f$ is Lipschitz continuous with constant
$2$:

\begin{align*}
|f(U_{1})-f(U_{2})| & \leq d_{\mathrm{HS}}(U_{1}C_{0}U_{1}^{\dagger},U_{2}C_{0}U_{2}^{\dagger})\\
 & \leq d_{\mathrm{HS}}(U_{1}C_{0}U_{1}^{\dagger},U_{1}C_{0}U_{2}^{\dagger}) +d_{\mathrm{HS}}(U_{1}C_{0}U_{2}^{\dagger},U_{2}C_{0}U_{2}^{\dagger})\\
 & =\|C_{0}(U_{1}^{\dagger}-U_{2}^{\dagger})\|_{\mathrm{HS}}+\|(U_{1}-U_{2})C_{0}\|_{\mathrm{HS}}\\
 & \leq2d_{\mathrm{HS}}(U_{1,}U_{2}).
\end{align*}
The generalized Levy's lemma \citep{measure_concentration1,measure_concentration2,Meckes2019}
states that for any Lipschitz continuous function over some Riemann
manifolds with positive curvature, its values are concentrated around
the mean one. For the unitary group, 
we have 
\begin{equation}
\mathbb{P}(|f(U)-\langle f(U)\rangle|\geq l\epsilon)\leq2e^{-\frac{N\epsilon^{2}}{12}},
\end{equation}
where $l$ is the Lipschitz constant. In what follows, we will bound $\langle f(U)\rangle=\langle d_{\mathrm{HS}}(\Pi_{A}UC_{0}U^{\dagger}\Pi_{A}^{\dagger},\langle C_{A}\rangle)\rangle$.
First, Since $\langle C_{A}\rangle=\Pi_{A}\int d_{\mathrm{H}}(U)UC_{0}U^{\dagger}\Pi_{A}^{\dagger}$
is invariant under any unitary on $\mathbb{U}(N_A)$, 
according to Schur's lemma, $\langle C_{A}\rangle=\frac{m}{N}I_{A}$
and
\begin{align*}
\langle\|\Pi_{A}UC_{0}U^{\dagger}\Pi_{A}^{\dagger}-\langle C_{A}\rangle\|_{\mathrm{HS}}\rangle & \leq\sqrt{\langle\|\Pi_{A}UC_{0}U^{\dagger}\Pi_{A}^{\dagger}-\langle C_{A}\rangle\|_{\mathrm{HS}}^{2}\rangle}\\
 & =\sqrt{\langle\mathrm{Tr}(\Pi_{A}UC_{0}U^{\dagger}\Pi_{A}^{\dagger}-\langle C_{A}\rangle)^{2}\rangle}\\
 & =\sqrt{\langle\mathrm{Tr}(\Pi_{A}UC_{0}U^{\dagger}\Pi_{A}^{\dagger})^{2}\rangle-\frac{m^{2}}{N^{2}}N_{A}}.
\end{align*}
Next, we need to calculate $\langle\Tr(\Pi_{A}UC_{0}U^{\dagger}\Pi_{A}^{\dagger})^{2}\rangle$.
The idea is similar as in \citep{Popescu2006} and originally comes
from random quantum channel coding \citep{PhysRevA.55.1613}: we introduce
another reference space $R'$ which has the same dimension as the
original total system $R$. The following equation holds:
\[
\langle\mathrm{Tr}(\Pi_{A}UC_{0}U^{\dagger}\Pi_{A}^{\dagger})^{2}\rangle=\int d_{H}(U)\mathrm{Tr}[(\Pi_{A}\otimes\Pi_{A})(UC_{0}U^{\dagger}\otimes UC_{0}U^{\dagger})\mathrm{SWAP}_{RR'}(\Pi_{A}^{\dagger}\otimes\Pi_{A}^{\dagger})],
\]
where $\mathrm{SWAP}_{RR'}$ is the SWAP operation between the original
system $R$ and the reference one $R'$. From Schur-Weyl duality \citep{Hayashi2017}, we obtain
\begin{equation}
\int d_{H}(U)(UC_{0}U^{\dagger}\otimes  UC_{0}U^{\dagger})\mathrm{SWAP}_{RR'}=\alpha I_{RR'}+\beta\mathrm{SWAP}_{RR'}.
\label{eq:unitary_2_design}
\end{equation}
Now, for simplicity, we can take $C_{0}=\begin{pmatrix}I_{m} & 0\\
0 & 0
\end{pmatrix}$. The following relations hold
\begin{align*}
\text{Tr}\mathrm{SWAP}_{RR'} & =N,\\
\mathrm{Tr}[(U\otimes  U)(C_{0}\otimes  C_{0}) (U^{\dagger}\otimes  U^{\dagger})\mathrm{SWAP}_{RR'}] & =\mathrm{Tr}[(U\otimes  U)(C_{0}\otimes  C_{0})\mathrm{SWAP}_{RR'} (U^{\dagger}\otimes  U^{\dagger})] =\mathrm{Tr}C_{0}^{2}=m,\\
\mathrm{Tr}[(U\otimes  U) (C_{0}\otimes  C_{0}) (U^{\dagger}\otimes  U^{\dagger})] & =m^{2}.
\end{align*}
The trace of Eq. (\ref{eq:unitary_2_design}) gives $N^{2}\alpha+N\beta=m$.
Multiplying Eq. (\ref{eq:unitary_2_design}) by $\mathrm{SWAP}_{RR'}$
and tracing it, we have $N\alpha+N^{2}\beta=m^{2}$. Solving the equations
leads to $\begin{cases}
\alpha= & \frac{Nm-m^{2}}{N(N^{2}-1)}\\
\beta= & \frac{Nm^{2}-m}{N(N^{2}-1)}
\end{cases}.$ As a result, ~
\begin{align}
\langle\mathrm{Tr}(\Pi_{A}UC_{0}U^{\dagger}\Pi_{A}^{\dagger})^{2}\rangle & =\mathrm{Tr}[(\Pi_{A}\otimes \Pi_{A})(\alpha I_{RR'}+\beta\mathrm{SWAP}_{RR'})(\Pi_{A}^{\dagger}\otimes \Pi_{A}^{\dagger}\nonumber)] \\
 & =\alpha N_{A}^{2}+\beta N_{A}.
 \label{eq:value_of_average_of_tracesquare}
\end{align}
Assuming that in the thermodynamic limit $N\to\infty$, the density of charge
$\frac{m}{N}$ is fixed as $\mathcal{O}(1)$, we obtain
\begin{align*}
\langle f(U)\rangle^{2} & \leq\frac{mN_{A}^{2}}{N(N-1)}\sim \mathcal{O}\left(\frac{N_{A}^{2}}{N}\right)
\end{align*}
and the typicality
\begin{equation}
\mathbb{P}\left(d_{\mathrm{HS}}\left(C_{A},\frac{m}{N}I_{A}\right)\geq2\epsilon+\sqrt{\frac{mN_{A}^{2}}{N(N-1)}}\right)\leq2e^{-\frac{N\epsilon^{2}}{12}}.
\label{eq:meassure_concentration_on_covariance_matrix}
\end{equation}

For the other direction, 
we take $f(U)=d_{\mathrm{HS}}^{2}(C_{A},\langle C_{A}\rangle)$,
which 
is also Lipschitz continuous with constant calculated as
\begin{align}
|f(U_{1})-f(U_{2})| & \leq2(d_{\mathrm{HS}}(\Pi_{A}U_{1}C_{0}U_{1}^{\dagger}\Pi_{A}^{\dagger},\langle C_{A}\rangle)+d_{\mathrm{HS}}(\Pi_{A}U_{2}C_{0}U_{2}^{\dagger}\Pi_{A}^{\dagger},\langle C_{A}\rangle))d_{\mathrm{HS}}(U_{1,}U_{2})\nonumber \\
 & \leq4\sqrt{N_{A}}\left(1-\frac{m}{N}\right)d_{\mathrm{HS}}(U_{1},U_{2}).
\end{align}
Here we assume $m\leq\frac{N}{2}$ due to the particle-hole symmetry (otherwise, we may replace $1-\frac{m}{N}$ by $\frac{m}{N}$). According to Eq. (\ref{eq:value_of_average_of_tracesquare}), we obtain
\begin{equation}
\langle f(U)\rangle=\langle\mathrm{Tr}(\Pi_{A}UC_{0}U^{\dagger}\Pi_{A}^{\dagger})^{2}\rangle-\frac{m^{2}}{N^{2}}N_{A}\geq\frac{(N-m)mN_{A}^{2}}{N^{2}(N+1)}.
\end{equation}
If $\frac{N_{A}}{N}$ and $\frac{m}{N}$ are both fixed as $\mathcal{O}(1)$ in the thermodynamic limit, this formula scales linear with $N$. Applying generalized
Levy's lemma leads to 
\begin{equation}
\mathbb{P}\left(d_{\mathrm{HS}}^{2}\left(C_{A},\frac{m}{N}I_{A}\right)\leq\frac{(N-m)mN_{A}^{2}}{N^{2}(N+1)}-4\sqrt{N_{A}}\left(1-\frac{m}{N}\right)\epsilon\right)\leq2e^{-\frac{N}{12}\epsilon^{2}}. 
\label{eq:atypical_expression}
\end{equation}
For example, if we choose $\epsilon\sim\mathcal{O}(N^{\frac{1}{3}})$,
the above inequaility means that $C_{A}$ will deviate from its
ensemble average by an $\mathcal{O}(N)$ factor with almost unit probability. 
This is the atypicality discussed in the main text.

\section{Some Applications of Measure Concentration Typicality}

\subsection{Measure concentration property for entropy}

In this subsection, we will use Eq. (\ref{eq:meassure_concentration_on_covariance_matrix})
and Eq. (\ref{eq:atypical_expression}) to derive the measure concentration
typicality/atypicality for subsystem entropy. For simplicity the half
filling condition is assumed. The eigenvalues of $C_{A}$ are denoted
as $\{\frac{1}{2}+\lambda_{i}\},i\in\{1,\cdots,N_{A}\}$ with $\sum_{i=1}^{N_{A}}\lambda_{i}^{2}=d_{\mathrm{HS}}(C_{A},\frac{I_{A}}{2})$
and $\lambda_{i}\in[-\frac{1}{2},\frac{1}{2}]$.

If the subsystem is microscopically small, we know the typicality of entropy
follows by noting that 
\begin{equation}
S_A=\sum_{i=1}^{N_{A}}H\left(\frac{1}{2}+\lambda_{i},\frac{1}{2}-\lambda_{i}\right)=\sum_{i=1}^{N_{A}}\left[1-\sum_{n=1}^{\infty}\frac{(2\lambda_{i})^{2n}}{2n(2n-1)\ln2}\right]\geq N_{A}-4d_{\mathrm{HS}}^{2}\left(C_{A},\frac{I_{A}}{2}\right),
\label{eq:Expansion_of_entropy_by_eigenvalues}
\end{equation}
where we replace $(2\lambda_i)^{2n}$ by $(2\lambda_i)^2$ in the last inequality since $(2\lambda_{i})^{2}\leq1$. Here $H(p_0,p_1)=-p_0\log_2 p_0 - p_1\log_2 p_1$ is the Shannon entropy. Combined with Eq. (\ref{eq:meassure_concentration_on_covariance_matrix})
we obtain
\begin{equation}
\begin{split}\mathbb{P}(N_{A}-S_A\geq x) & \leq\mathbb{P}\left(d_{\mathrm{HS}}\left(C_{A},\frac{I_{A}}{2}\right)\geq\frac{\sqrt{x}}{2}\right)\\
 & \leq\begin{cases}
2\exp\left[-\frac{N}{48}\left(\frac{\sqrt{x}}{2}-\sqrt{\frac{N_{A}^{2}}{2(N-1)}}\right)^{2}\right], & x>\frac{2N_{A}^{2}}{N-1};\\
1, & x\leq\frac{2N_{A}^{2}}{N-1}.
\end{cases}
\label{eq:typicality_for_subsystem_entropy}
\end{split}
\end{equation}
As long as $N\gg N_{A}^{2}$, we conclude the subsystem entropy will
be nearly maximal. 

For the other direction, if $N_{A}$ is macroscopically large, we can upper 
bound the lhs of Eq. (\ref{eq:Expansion_of_entropy_by_eigenvalues}) by
\begin{equation}
S_A\leq N_{A}-\frac{2}{\ln2}d_{\mathrm{HS}}^{2}\left(C_{A},\frac{I_{A}}{2}\right).
\end{equation}
Following the atypicality of $d_{\mathrm{HS}}^{2}(C_{A},\frac{I_{A}}{2})$
in Eq.~(\ref{eq:atypical_expression}), the subsystem entropy density
will show an $\mathcal{O}(1)$ deviation from the maximal value
:
\begin{equation}
\begin{split}\mathbb{P}\left(S_A\geq N_{A}-\frac{N_{A}^{2}}{2\ln2(N+1)}+\frac{2}{\ln2}\epsilon\right) & \leq\mathbb{P}\left(d_{\mathrm{HS}}^{2}\left(C_{A},\frac{I_{A}}{2}\right)\leq\frac{N_{A}^{2}}{4(N+1)}-\epsilon\right)\\
 & \leq2e^{-\frac{N}{48N_{A}}\epsilon^{2}}.
\end{split}
\end{equation}
We may take $\epsilon\sim\mathcal{O}(N^{\alpha})$ for arbitrary $\alpha\in(0,1)$,
finding that the majority of subsystem entropy will be comparable
or smaller than $N_{A}-\frac{N_{A}^{2}}{2\ln2(N+1)}$. This clearly
illustrates the difference of the Page curve for the RFG ensemble from
the interacting one.

\subsection{Upper bound on the variance of entropy}

At the end of this section, we will discuss the variance of entropy
in microscopic region. Here the half filling condition is also assumed.
From Eq. (\ref{eq:typicality_for_subsystem_entropy}), we obtain
\[
\mathbb{P}((S_A-N_{A})^{2}\geq x)\leq\begin{cases}
1, & x\leq\frac{4N_{A}^{4}}{(N-1)^{2}};\\
2\exp\left[-\frac{N}{48}\left(\frac{x^{\frac{1}{4}}}{2}-\sqrt{\frac{N_{A}^{2}}{2(N-1)}}\right)^{2}\right], & x>\frac{4N_{A}^{4}}{(N-1)^{2}}.
\end{cases}
\]
Therefore
\begin{align*}
\langle(S_A-N_{A})^{2}\rangle & =\int\mathbb{P}((S_A-N_{A})^{2}\geq x)dx\\
 & \leq\frac{4N_{A}^{4}}{(N-1)^{2}}+2\int_{\frac{4N_{A}^{4}}{(N-1)^{2}}}^{\infty}dx\exp\left[-\frac{N}{48}\left(\frac{x^{\frac{1}{4}}}{2}-\sqrt{\frac{N_{A}^{2}}{2(N-1)}}\right)^{2}\right].
\end{align*}
For the last line, we can change the integral variable into $t=\sqrt{N}\left(\frac{x^{\frac{1}{4}}}{2}-\sqrt{\frac{N_A^2}{2(N-1)}}\right)$, obtaining
\begin{align*}
 & 2\int_{\frac{4N_{A}^{4}}{(N-1)^{2}}}^{\infty}dx\exp\left[-\frac{N}{48}\left(\frac{x^{\frac{1}{4}}}{2}-\sqrt{\frac{N_{A}^{2}}{2(N-1)}}\right)^{2}\right]
=  \frac{128}{\sqrt{N}}\int_{0}^{\infty}dt\left(\frac{t}{\sqrt{N}}+\sqrt{\frac{N_{A}^{2}}{2(N-1)}}\right){}^{3}e^{-\frac{t^{2}}{48}}\\
= & \frac{128}{N^{2}}\int_{0}^{\infty}dt\left(t+\sqrt{\frac{NN_{A}^{2}}{2(N-1)}}\right)^{3}e^{-\frac{t^{2}}{48}}
\sim  \mathcal{O}\left(\frac{1}{N^{2}}\right),
\end{align*}
provided that $N_{A}$ is fixed as $\mathcal{O}(1)$. In conclusion, we obtain
\[
\mathrm{Var}(S_A)\leq\langle(S_A-N_{A})^{2}\rangle\sim\mathcal{O}(N^{-2}), 
\]
which agrees with \citep{Bianchi2021,Bianchi2021a}.

\section{Detailed Calculation of the Dynamical 
Page Curves}
As mentioned in the main text, for all the models in this section, we assume the initial state is a period-2 density wave with half filling. Following the same notation in the main text, we further define $X_{A}(t)=2C_{A}(t)-I_{A}$. Thus
\begin{equation}
S_{A}(t)=N_{A}-\sum_{n=1}^{\infty}\frac{\Tr X_{A}^{2n}(t)}{2n(2n-1)\ln2}.
\label{eq:Expanding_of_entropy_with_X}
\end{equation}

\subsection{Calculation for the minimal model}
We first consider the minimal model. 
Remember that the minimal model means only nearest neighbor hopping
is included. 
 After introducing the Fourier transformed 
 mode $a_{k}^{\dagger}=\frac{1}{\sqrt{N}}\sum_{j=1}^{N}e^{-ikj}a_{j}^{\dagger}$,
we can easily obtain the correlation function in momentum space
\begin{equation}
\Tr[\rho a_{k}^{\dagger}(t)a_{k'}(t)]=\frac{1}{2}\delta_{k,k'}+\frac{1}{2}\delta_{k,k'+\pi}e^{i\theta_{k}(t)},
\end{equation}
where $a_k(t)$ ($a^\dag_{k'}(t)$) is the annihilation (creation) operator in the Heisenberg picture, $\theta_{k}(t)=t(E_{k}-E_{k+\pi})$ and $\rho=|\Psi_0\rangle\langle\Psi_0|$ corresponds
to the initial density matrix. In the following, we may omit 
the
index $t$ if there is no ambiguity.

After the inverse Fourier transformation back to the position space, the covariance
matrix for subsystem $A$ reads  $[C_A]_{m_{1}m_{2}}=\frac{\delta_{m_{1},m_{2}}}{2}+\frac{1}{2N}\sum_{k}e^{i\theta_{k}}e^{ik(m_{1}-m_{2})}e^{i\pi m_{2}}$
and thus
\begin{equation}
[X_A]_{m_{1}m_{2}}=\frac{1}{N}\sum_{k}e^{i\theta_{k}}e^{ik(m_{1}-m_{2})}e^{i\pi m_{2}}.
\label{eq:Expression_for_X_in_NNH}
\end{equation}

\subsubsection{Second order in $X_{A}$}

With Eq. (\ref{eq:Expression_for_X_in_NNH}), we can calculate $\overline{\Tr X_{A}^{2}}$ as
\begin{align*}
\overline{\Tr X_{A}^{2}} & =\frac{1}{N^{2}}\sum_{k_{1},k_{2},m_{1},m_{2}}\overline{e^{i\theta_{k_{1}}}e^{ik_{1}(m_{1}-m_{2})}e^{i\pi m_{2}}e^{i\theta_{k_{2}}}e^{ik_{2}(m_{2}-m_{1})}e^{i\pi m_{1}}}\\
 & =\frac{1}{N^{2}}\sum_{k_{1},k_{2},m_{1},m_{2}}(\delta_{k_{1},k_{2}+\pi}e^{ik_{1}(m_{1}-m_{2})}e^{ik_{1}(m_{2}-m_{1})}e^{-i\pi(m_{2}-m_{1})}e^{i\pi(m_{1}+m_{2})}\\
 & +\delta_{k_{1}+k_{2},\pi}e^{i2k_{1}(m_{1}-m_{2})}e^{i\pi(m_{1}+m_{2})}e^{i\pi(m_{2}-m_{1})})\\
 & =\frac{N_{A}^{2}}{N}+\frac{1}{N}\sum_{k,m_{1},m_{2}}e^{i2k(m_{1}-m_{2})}=\frac{N_{A}^{2}}{N}+\frac{1}{N}\sum_{m_{1},m_{2}}(\delta_{m_{1}-m_{2},0}+\delta_{m_{1}-m_{2},\frac{N}{2}}+\delta_{m_{1}-m_{2},-\frac{N}{2}}).
\end{align*}
Noting that the first term in the middle step comes from $\theta_{k+\pi}=-\theta_{k}$,
which holds for general Hamiltonians according to the definition. However,
the second term is due to the reflection symmetry (which implies $E_k=E_{-k}$) of 
the minimal model and is thus not universal. 
Fortunately, this model dependent term will vanish in the thermodynamic limit.

\begin{figure*}
\includegraphics[width=1\textwidth]{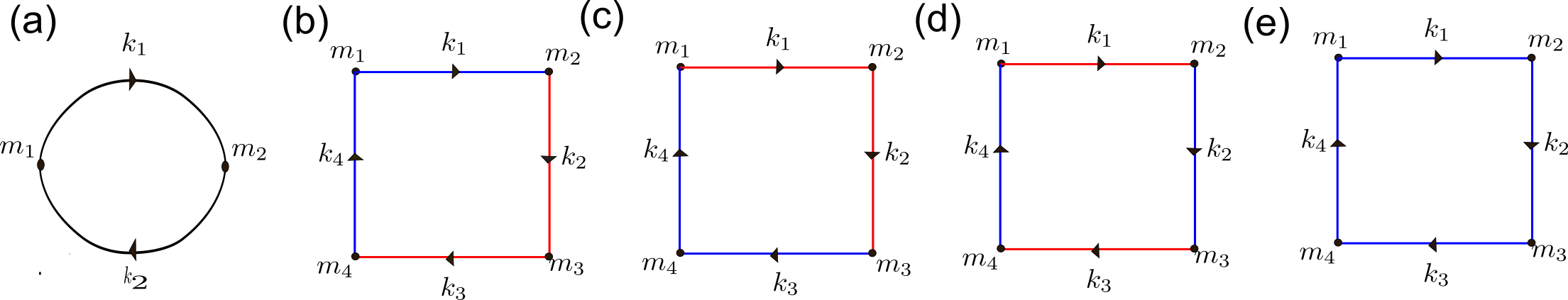}
\caption{
Feynman diagrams for calculating the entanglement entropy order by order. Here each
vertex represents a position index and each leg represents a momentum
index. Each leg is associated with $
e^{i\theta_{k}}e^{ik(m-m')}e^{i\pi m'}$.
In these diagrams, the legs with same color need to be contracted.
Each color corresponds to one contraction.}
\label{Total_figure_for_FD_1}
\end{figure*}

This calculation can be diagrammatically represented as shown in Fig.~\ref{Total_figure_for_FD_1}(a), where we draw an arrow from $m_1$ to $m_2$ ($m_2$ to $m_1$) because there is a factor $e^{ik_1(m_1-m_2)}$ ($e^{ik_2(m_2-m_1)}$). As will become clear below, such a diagrammatic representation provides a convenient and systematic way for dealing with higher order terms.

\subsubsection{Third order in $X_{A}$}
We move on to calculate $\overline{\Tr X_{A}^{3}}$ and will 
see 
it vanishes in thermodynamic limit. The expression is 
\begin{align*}
\overline{\Tr X_{A}^{3}} & =\frac{1}{N^{3}}\sum_{k_{1,2,3}m_{1,2,3}}\overline{e^{i\theta_{k_{1}}+i\theta_{k_{2}}+i\theta_{k_{3}}}}e^{ik_{1}(m_{1}-m_{2})+ik_{2}(m_{2}-m_{3})+ik_{3}(m_{3}-m_{1})}e^{i\pi(m_{1}+m_{2}+m_{3})}.
\end{align*}

The non-zero contribution of $\overline{e^{i\theta_{k_{1}}+i\theta_{k_{2}}+i\theta_{k_{3}}}}$
in the minimal model only comes from two cases:
\begin{enumerate}
\item $k_{2}=k_{1}+\pi,k_{3}=\pm\frac{\pi}{2}$ and cyclic permutations.
The contribution is proportional to 
\begin{align*}
 & \frac{1}{N^{3}}\sum_{k_{1},m_{1},m_{2},m_{3}}e^{ik_{1}(m_{1}-m_{3})}e^{i\frac{\pi}{2}(m_{1}+m_{3})} =\frac{N_{A}}{N^{2}}\sum_{m_{1},m_{3}}\delta_{m_{1},m_{3}}e^{i\pi m_{1}}\sim \mathcal{O}\left(\frac{1}{N}\right).
\end{align*}
\item $k_{1},k_{2}$ satisfy $|\cos k_{1}+\cos k_{2}|\leq1$ and $k_{3}=\arccos(-\cos k_{1}-\cos k_{2})$.
The contribution is proportional to
\[
\sim\frac{1}{N^{3}}\sum_{k_{1},k_{2}}\sum_{m_{1}}e^{im_{1}(k_{1}-k_{3}+\pi)}\sum_{m_{2}}e^{im_{2}(k_{2}-k_{1}+\pi)}\sum_{m_{3}}e^{im_{3}(k_{3}-k_{2}+\pi)}.
\]
With H\"older inequality \citep{hardy1988inequalities}: $\sum_{i}|a_{i}||b_{i}||c_{i}|\leq[(\sum_{i}|a_{i}|^{3})(\sum_{i}|b_{i}|^{3})(\sum_{i}|c_{i}|^{3})]^{\frac{1}{3}}$,
we can upper bound the above contribution as 
\[
\leq\frac{1}{N^{3}}\{[\sum_{k_{1},k_{2}}|\sum_{m_{1}}e^{im_{1}(k_{1}-k_{3}+\pi)}|^{3}][\sum_{k_{1},k_{2}}|\sum_{m_{2}}e^{im_{2}(k_{2}-k_{1}+\pi)}|^{3}][\sum_{k_{3},k_{2}}|\sum_{m_{3}}e^{im_{3}(k_{3}-k_{2}+\pi)}|^{3}]\}^{\frac{1}{3}}.
\]
Since here, no pair of two $k$ differ by $\pi$ (as it is the case
already considered in the first case), the sum of $m_{1},m_{2},m_{3}$
only contributes to $\mathcal{O}(1)$, so the total contribution will be upper
bounded by $\mathcal{O}\left(\frac{1}{N}\right)$.
\end{enumerate}
In conclusion
\[
\overline{\Tr X_{A}^{3}}\sim \mathcal{O}\left(\frac{1}{N}\right)
\]
and thus vanishes in the thermodynamic limit. 

The above discussion can be generalized to higher orders: as long
as the degeneracy point of a Hamiltonian is not dense, we can safely
ignore the model dependent contribution 
and put $\overline{e^{i\theta_{k_{1}}}e^{i\theta_{k_{2}}}}=\delta_{k_{1},k_{2}+\pi}$,
which we call 
a contraction. In the following, 
we will directly
use this contraction rule.\footnote{Another illustration for general cases can be obtained with the techniques in Subsec.~\ref{subsec:Proof-of-main}.}

\subsubsection{Forth Order in $X_{A}$}

Now we are moving to calculate $\overline{\Tr X_{A}^{4}}$:
\begin{equation}
\overline{\Tr X_{A}^{4}}=\frac{1}{N^{4}}\sum_{k_{1,2,3,4},m_{1,2,3,4}}\overline{\prod_{j}^{4}e^{i\theta_{k_{j}}}e^{ik_{j}(m_{j}-m_{j+1})}e^{im_{j}\pi}},
\label{eq:Expression_for_XA_4}
\end{equation}
where $m_{5}=m_{1}$. 

The first contracting class for $\overline{e^{i\theta_{k_{1}}}e^{i\theta_{k_{2}}}e^{i\theta_{k_{3}}}e^{i\theta_{k_{4}}}}$
is to contract the dynamical phase factors in pairs. 
Two patterns in this class are shown
in Fig.~\ref{Total_figure_for_FD_1}(b) and Fig.~\ref{Total_figure_for_FD_1}(c).
The legs with same colors mean that they are contracted together.
These patterns correspond to $\overline{e^{i\theta_{k_{1}}}e^{i\theta_{k_{4}}}}\;\overline{e^{i\theta_{k_{2}}}e^{i\theta_{k_{3}}}}=\delta_{k_{1},k_{4}+\pi}\delta_{k_{2},k_{3}+\pi}$
and $\overline{e^{i\theta_{k_{1}}}e^{i\theta_{k_{2}}}}\;\overline{e^{i\theta_{k_{4}}}e^{i\theta_{k_{3}}}}=\delta_{k_{1},k_{2}+\pi}\delta_{k_{4},k_{3}+\pi}$,
respectively. Substituting these delta functions into Eq. (\ref{eq:Expression_for_XA_4}),
we obtain 
\begin{equation}
2\times\frac{N_{A}^{2}}{N^{4}}\sum_{k_{1,3},m_{1,3}}e^{ik_{1}(m_{1}-m_{3})}e^{ik_{3}(m_{3}-m_{1})}=\frac{2N_{A}^{3}}{N^{2}},
\end{equation}
where the factor $2$ comes from the equal contribution of these two diagrams. 

Another pattern from this contracting class is shown in Fig.~\ref{Total_figure_for_FD_1}(d)
which gives $\overline{e^{i\theta_{k_{1}}}e^{i\theta_{k_{3}}}}\;\overline{e^{i\theta_{k_{2}}}e^{i\theta_{k_{4}}}}=\delta_{k_{1},k_{3}+\pi}\delta_{k_{2},k_{4}+\pi}$.
However, substituting this expression into Eq. (\ref{eq:Expression_for_XA_4})
leads to
\begin{equation}
\begin{aligned}\frac{1}{N^{4}}\sum_{k_{1,2},m_{1,2,3,4}}e^{i(k_{1}-k_{2})(m_{1}+m_{3}-m_{2}-m_{4})}e^{i\pi(m_{2}+m_{4})} & =\frac{1}{N^{2}}\sum_{m_{1,2,3,4}}\delta_{m_{1}+m_{3},m_{2}+m_{4}}e^{i\pi(m_{2}+m_{4})} \sim \mathcal{O}\left(\frac{1}{N}\right).
\end{aligned}
\end{equation}
This diagram thus 
vanishes in the thermodynamic limit. 

It should be emphasized that, in the above discussion, some terms 
are calculated multiple times.
These form the other contracting class: we contract the four legs
all together, as shown in Fig.~\ref{Total_figure_for_FD_1}(e). The
contribution from this diagram needs to be subtracted due to the multiple
calculation in Fig.~\ref{Total_figure_for_FD_1}(b) and (c):
\begin{equation}
-\frac{1}{N^{4}}\sum_{k_{1,2,3,4},m_{1,2,3,4}}\delta_{k_{2},k_{1}+\pi}\delta_{k_{3},k_{1}}\delta_{k_{4},k_{1}+\pi}\prod_{j}^{4}e^{ik_{j}(m_{j}-m_{j+1})}e^{im_{j}\pi}=-\frac{N_{A}^{4}}{N^{3}}.
\end{equation}
Combining all the contributions 
together, we obtain 
\begin{equation}
\overline{\Tr X_{A}^{4}}=\frac{2N_{A}^{3}}{N^{2}}-\frac{N_{A}^{4}}{N^{3}}.
\end{equation}

From the above discussion, we can see that the contraction rules here
are obviously different from 
Wick's theorem.

\subsection{General Feynman Rules}

\begin{figure*}
\includegraphics[width=0.8\textwidth]{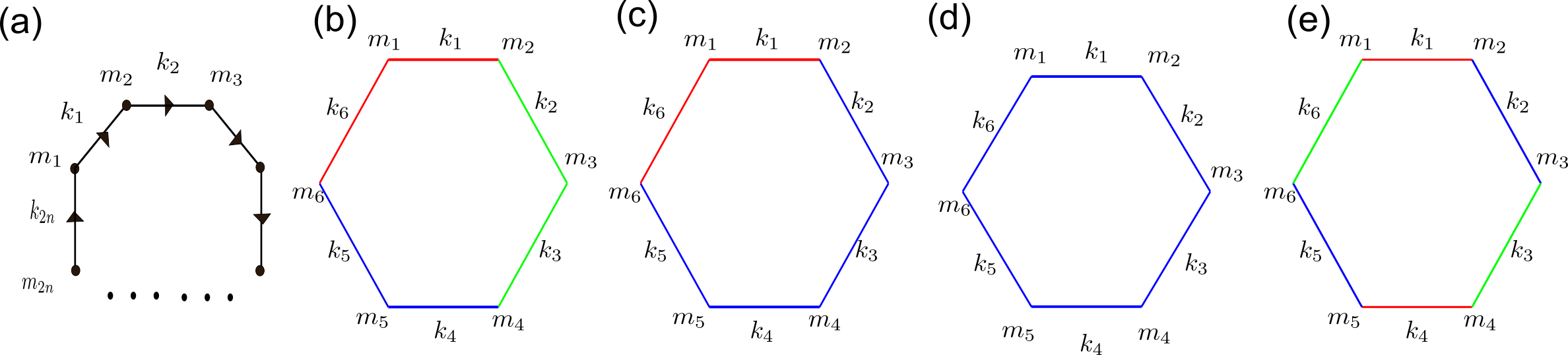}
\caption{In this figure, more complicated Feynman diagrams are shown compared 
to Fig.~\ref{Total_figure_for_FD_1}. (a) shows a general Feynman diagram,
(b)-(e) are the Feynman diagrams for calculating $\overline{\Tr X_{A}^{6}}$.}
\label{Total_figure_for_FD_2}
\end{figure*}


The method presented in the previous subsection allows us to calculate
Eq. (\ref{eq:Expanding_of_entropy_with_X}) to arbitrary orders. Here
we summarize our Feynman rules for contraction. A general Feynman diagram
for calculating $\overline{\Tr X_{A}^{2n}}$ is shown in Fig. \ref{Total_figure_for_FD_2}(a).
We will use the integer $i$ to label the legs associated with momentum
$k_{i}$. 
\begin{enumerate}
\item All the legs in Fig. \ref{Total_figure_for_FD_2}(a) must be contracted.  Each contraction leads to a delta function of momenta $k$s and has to include even number of legs, where half of the legs should be labeled
as even and the other half as odd. This last requirement arises from the phase factor $e^{i\pi m}$ in each leg and is to ensure the diagram 
not to vanish in the thermodynamic limit 
(we recall that Fig.~\ref{Total_figure_for_FD_1}(d) does not contribute as this requirement is not satisfied). For totally $2n$ legs with $l$ contractions, the $N,N_{A}$ dependence for this diagram is
$\frac{N_{A}^{2n-l+1}}{N^{2n-l}}$.
\item Each contraction with $2j$ legs should also be assigned a multiple
factor $a_{2j}$, accounting for the multiple calculations. $a_{2}$ and $a_{4}$ are obtained in previous discussion while higher $a_{2j}$ can be obtained iteratively, as shown below.
\item For each diagram, multiply the term $\frac{N_{A}^{2n-l+1}}{N^{2n-l}}$ and factors $a_{2j}$ obtained in Rule 1 and Rule 2 together. Some diagrams also need to multiply by subsystem correction factor $\beta$ (see details below). Sum over all possible diagrams leads to the desired result.
\end{enumerate}
The subsystem correction factor $\beta$ does not appear in $\overline{\Tr X_{A}^{4}}$ and $\overline{\Tr X_{A}^{2}}$, but will appear in calculating $\overline{\Tr X_{A}^{6}}$. For pedagogical purpose, we will now show how to calculate $\overline{\Tr X_{A}^{6}}$. Some diagrams are shown in Fig. \ref{Total_figure_for_FD_2}(b-e).
In diagram (b), there are three contractions, each contracts two legs.
The contribution for this diagram is $\frac{N_{A}^{4}}{N^{3}}a_{2}^{3}$.
In diagram (c), there are two contractions, one contracts four legs
together and the other contracts two legs. The contribution is $\frac{N_{A}^{5}}{N^{4}}a_{4}a_{2}$.
In diagram (d), all the six legs are contracted together, contributing
to $\frac{N_{A}^{6}}{N^{5}}a_{6}$. Attentions should be payed to 
diagram (e). After contracting the three pairs of legs, we obtain
\begin{equation}
\begin{split} & \sum_{k_{1,2,3}m_{1,2,3,4,5,6}}e^{ik_{1}(m_{1}-m_{2}+m_{4}-m_{5})}e^{ik_{2}(m_{2}-m_{3}+m_{5}-m_{6})}e^{ik_{3}(m_{3}-m_{4}+m_{6}-m_{1})}\\
= & N^{3}\sum_{m_{1,2,3,4,5,6}}\delta_{m_{1}+m_{4}=m_{2}+m_{5}=m_{3}+m_{6}\ \mathrm{mod\ }N}.
\end{split}
\end{equation}
Naively, one may conjecture 
the result of the last sum to 
be $N_{A}^{4}$.
However, this is only true when $f=\frac{N_{A}}{N}=1$. If $f\leq\frac{1}{2}$,
the sum will be much smaller. After carefully counting the pairs satisfying
the delta function, we obtain for $f\leq\frac{1}{2}$
\begin{equation}
\sum_{k_{1,2,3}m_{1,2,3,4,5,6}}e^{ik_{1}(m_{1}-m_{2}+m_{4}-m_{5})}e^{ik_{2}(m_{2}-m_{3}+m_{5}-m_{6})}e^{ik_{3}(m_{3}-m_{4}+m_{6}-m_{1})}=\frac{1}{2}N^{3}(N_{A}^{4}+N_{A}^{2})\to\beta_{1}N^{3}N_{A}^{4},
\end{equation}
where $\beta_{1}$ is defined as the subsystem correction factor in
thermodynamic limit: $\beta_{1}=\begin{cases}
\frac{1}{2} & f\leq\frac{1}{2}\\
1 & f=1
\end{cases}$. Summing over all possible diagrams, the total contribution is 
\begin{equation}
\overline{\Tr X_{A}^{6}}=(5+\beta_{1})\frac{N_{A}^{4}}{N^{3}}-(6+3\beta_{2})\frac{N_{A}^{5}}{N^{4}}+\frac{N_{A}^{6}}{N^{5}}a_{6},
\end{equation}
where $\beta_{2}$ is another subsystem correction factor $\beta_{2}=\begin{cases}
\frac{2}{3} & f\leq\frac{1}{2}\\
1 & f=1
\end{cases}$. Here $a_{6}$ can be determined by considering the case when $N_{A}=N$
(i.e. $f=1$). In this case, $\overline{\Tr X_{A}^{6}}=N$, resulting
in $a_{6}=4$. Therefore, 
\begin{equation}
\overline{\Tr X_{A}^{6}}=\frac{11}{2}\frac{N_{A}^{4}}{N^{3}}-8\frac{N_{A}^{5}}{N^{4}}+4\frac{N_{A}^{6}}{N^{5}}\ \ \mathrm{if}\ \ f\leq\frac{1}{2}.
\end{equation}
Following the above procedure for calculating the Feynman diagrams, we arrive at, up to the order $\mathcal{O}(f^5)$,
\begin{equation}
\frac{\overline{S_{A}}}{N}=f-\frac{1}{\ln{2}}\left(\frac{1}{2}f^{2}+\frac{1}{6}f^{3}+\frac{1}{10}f^{4}+0.06f^{5}\right)+\mathcal{O}(f^{6}).
\end{equation}

\subsection{Proof of 
Theorem 2 \label{subsec:Proof-of-main}}

In this subsection, we go 
beyond the minimal model and consider the general Hamiltonians satisfying the condition in 
Theorem 2 in the main text. We will find the Feynman rule as well as the subsystem entropy
is indeed the same as the previous subsection.

Since the Hamiltonian is period-2, we can use a modified Fourier
transformation to block diagonalize it:
\begin{equation}
A_{k}^{\dagger}=\sqrt{\frac{2}{N}}\sum_{j=1}^{\frac{N}{2}}e^{-ik(2j-1)}a_{2j-1}^{\dagger},\ \ B_k^\dag=\sqrt{\frac{2}{N}}\sum_{j=1}^{\frac{N}{2}}e^{-i2kj}a_{2j}^{\dagger},\;\;
k\in\left\{\frac{2n\pi}{N}\right\}^{\frac{N}{2}-1}_{n=0}.
\label{eq:FT_2_periodic}
\end{equation}
Here $A_{k}^{\dagger},B_{k}^{\dagger}$ are related to the conserved (eigen) modes
$P_{k}^{\dagger},Q_{k}^{\dagger}$ via a $2\times2$ unitary transformation
$U^{k}$as $A_{k}^{\dagger}=U_{11}^{k}P_{k}^{\dagger}+U_{12}^{k}Q_{k}^{\dagger}$
and $B_{k}^{\dagger}=U_{21}^{k}P_{k}^{\dagger}+U_{22}^{k}Q_{k}^{\dagger}$.
Substituting into Eq. (\ref{eq:FT_2_periodic}) leads to 
\begin{align}
a_{2m}^{\dagger} & =\sqrt{\frac{2}{N}}\sum_{k=0}^{\pi}e^{ik2m}(U_{21}^{k}P_{k}^{\dagger}+U_{22}^{k}Q_{k}^{\dagger}),\;\;\;\;
a_{2m+1}^{\dagger}  =\sqrt{\frac{2}{N}}\sum_{k=0}^{\pi}e^{ik(2m+1)}(U_{11}^{k}P_{k}^{\dagger}+U_{12}^{k}Q_{k}^{\dagger}).
\label{eq:IFT}
\end{align}
If we define $Q_{k+\pi}=P_{k}$ 
and 
\begin{equation}
Z_{k}^{m}=\begin{cases}
\sqrt{2}U_{22}^{k}, & \mathrm{if\ }m\ \mathrm{is\ even\ and}\ k<\pi;\\
\sqrt{2}U_{21}^{k-\pi}, & \mathrm{if\ }m\ \mathrm{is\ even\ and}\ k\geq\pi;\\
\sqrt{2}U_{12}^{k}, & \mathrm{if\ }m\ \mathrm{is\ odd\ and}\ k<\pi;\\
-\sqrt{2}U_{11}^{k-\pi}, & \mathrm{if\ }m\ \mathrm{is\ odd\ and}\ k\geq\pi,
\end{cases}
\label{eq:Zkm}
\end{equation}
the above inverse Fourier transformation (\ref{eq:IFT}) can be rewritten as
\[
a_{m}^{\dagger}=\frac{1}{\sqrt{N}}\sum_{k=0}^{2\pi}Z_{k}^{m}Q_{k}^{\dagger}e^{ikm}.
\]
In the following, we will simplify 
$\sum_{k=0}^{2\pi}$ as $\sum_{k}$
and $k$ should be understood as module $2\pi$. By assumption, all
the conserved quantity $\Tr(\rho Q_{k}^{\dagger}Q_{k})=\frac{1}{2}$
for $k\in[0,2\pi)$, thus
\begin{align*}
[C_A]_{ml} & =\frac{1}{2N}\sum_{k}Z_{k}^{m}Z_{k}^{l*}e^{ik(m-l)}+\frac{1}{2N}\sum_{k}e^{i\theta_{k}}Z_{k}^{m}Z_{k+\pi}^{l*}e^{ik(m-l)}e^{i\pi l}\\
 & =\frac{1}{2}\delta_{m,l}+\frac{1}{2N}\sum_{k}e^{i\theta_{k}}Z_{k}^{m}Z_{k+\pi}^{l*}e^{ik(m-l)}e^{i\pi l}
\end{align*}
where $\theta_{k+\pi}=-\theta_{k}$ by definition. In the last equality, we have
used the unitarity of $U^{k}$. Namely, if $m-l$ is odd:
\begin{align*}
\frac{1}{2N}\sum_{k}Z_{k}^{m}Z_{k}^{l*}e^{ik(m-l)} & =\frac{1}{4N}\sum_{k}[Z_{k}^{m}Z_{k}^{l*}e^{ik(m-l)}+Z_{k+\pi}^{m}Z_{k+\pi}^{l*}e^{i(k+\pi)(m-l)}]\\
 & =\frac{1}{4N}\sum_{k}(Z_{k}^{m}Z_{k}^{l*}-Z_{k+\pi}^{m}Z_{k+\pi}^{l*})e^{ik(m-l)}=0.
\end{align*}
A similar calculation can be carried out for the case in which 
$m-l$ is even. Therefore, we 
have 
\[
[X_A]_{ml}=\frac{1}{N}\sum_{k}e^{i\theta_{k}}Z_{k}^{m}Z_{k+\pi}^{l*}e^{ik(m-l)}e^{i\pi l}.
\]
This expression is very similar to Eq. (\ref{eq:Expression_for_X_in_NNH})
except for the extra factors 
$Z_{k}$. Nonetheless, we will show those extra
$Z_{k}$'s do 
not contribute 
in the thermodynamic limit. As a result,
the same Feynman rules and dynamical Page curve 
follows. 

When evaluating a Feynman diagram in the thermodynamic limit with 
$N_{A}$, $N$ both going to infinity,
we can first sum over the position indices. Introducing $s_{j}=k_{j}-k_{j-1}$, $k_{0}=k_{2n}$
and $m_{2n+1}=m_{1}$, we obtain
\begin{align*}
 & \sum_{m_{1,2\cdots2n}}\prod_{j=1}^{2n}e^{ik_{j}(m_{j}-m_{j+1})}e^{i\pi m_{j}}Z_{k_{j}}^{m_{j}}Z_{k_{j-1}+\pi}^{m_{j}*}\\
 & =\sum_{m_{1,2\cdots2n}}\prod_{j=1}^{2n}e^{im_{j}s_{j}}e^{i\pi m_{j}}Z_{k_{j}}^{m_{j}}Z_{k_{j-1}+\pi}^{m_{j}*}\\
 & =\prod_{j=1}^{2n}\left[\sum_{m_{j}:\mathrm{even}}e^{im_{j}(s_{j}+\pi)}Z_{k_{j}}^{0}Z_{k_{j-1}+\pi}^{0*}+\sum_{m_{j}:\mathrm{odd}}e^{im_{j}(s_{j}+\pi)}Z_{k_{j}}^{1}Z_{k_{j-1}+\pi}^{1*}\right]\\
 & =\prod_{j=1}^{2n}\left[\frac{1-e^{iN_{A}(s_{j}+\pi)}}{1-e^{2i(s_{j}+\pi)}}\right]\prod_{j=1}^{2n}(Z_{k_{j}}^{0}Z_{k_{j-1}+\pi}^{0*}+e^{i(s_{j}+\pi)}Z_{k_{j}}^{1}Z_{k_{j-1}+\pi}^{1*})\\
 & =\frac{e^{i\frac{N_{A}}{2}\sum_{j=1}^{2n}(s_{j}+\pi)}}{e^{i\sum_{j=1}^{2n}(s_{j}+\pi)}}\prod_{j=1}^{2n}\left[\frac{\sin\frac{N_{A}(s_{j}+\pi)}{2}}{\sin(s_{j}+\pi)}\right]\prod_{j=1}^{2n}(Z_{k_{j}}^{0}Z_{k_{j-1}+\pi}^{0*}+e^{i(s_{j}+\pi)}Z_{k_{j}}^{1}Z_{k_{j-1}+\pi}^{1*})\\
 & =\prod_{j=1}^{2n}\left[\frac{\sin\frac{N_{A}(s_{j}+\pi)}{2}}{\sin(s_{j}+\pi)}\right]\prod_{j=1}^{2n}(Z_{k_{j}}^{0}Z_{k_{j-1}+\pi}^{0*}+e^{i(s_{j}+\pi)}Z_{k_{j}}^{1}Z_{k_{j-1}+\pi}^{1*})
\end{align*}
In the above calculation, we have assumed $N_{A}$ to be even for simplicity. We also emphasize that $s_{1}\cdots s_{2n}$ is not independent since $\sum_{i=1}^{2n}s_{i}=0$.
The factor $\frac{\sin\frac{N_{A}(s_{j}+\pi)}{2}}{\sin(s_{j}+\pi)}$
will be dominated by the contribution from $s_{j}=0$ or $s_{j}=\pi$
if $Z_{k}$ is smooth enough. This is due to the convergence of Fourier
series, see \citep{stein2003fourier}. In \citep{Kress1998}
the difference between the discrete sum over momenta and the integration is also upper bounded. However, if $s_{j}\simeq
0$, it will lead to 
\begin{equation}
Z_{k_{j}}^{0}Z_{k_{j-1}+\pi}^{0*}+e^{i(s_{j}+\pi)}Z_{k_{j}}^{1}Z_{k_{j-1}+\pi}^{1*}\simeq
Z_{k_{j}}^{0}Z_{k_{j}+\pi}^{0*}-Z_{k_{j}}^{1}Z_{k_{j}+\pi}^{1*}=0
\end{equation}
due to the unitarity of $U$. In the end, 
we obtain 
\begin{align}
\sum_{m_{1,2\cdots2n}}\prod_{j=1}^{2n}e^{ik_{j}(m_{j}-m_{j+1})}e^{i\pi m_{j}}Z_{k_{j}}^{m_{j}}Z_{k_{j-1}+\pi}^{m_{j}*} & \simeq(Z_{k}^{0}Z_{k}^{0*}+Z_{k}^{1}Z_{k}^{1*})^{2n}\prod_{j=1}^{2n}\left[\frac{\sin\frac{N_{A}(s_{j}+\pi)}{2}}{\sin(s_{j}+\pi)}\right]\,s_{j}\neq 0\nonumber \\
 & =2^{2n}\prod_{j=1}^{2n}\left[\frac{\sin\frac{N_{A}(s_{j}+\pi)}{2}}{\sin(s_{j}+\pi)}\right]\ s_{j}\neq 0
 \label{eq:Summing_Over_position}
\end{align}
 Now we can see in the final expression Eq. (\ref{eq:Summing_Over_position}) that the model-dependent factor 
$Z$ disappears. Therefore, those Hamiltonians
satisfying the conditions in 
Theorem 2 in the main text will have the same Feynman
rules and dynamical Page curve 
as the minimal model.

\begin{figure}
\begin{center}
\includegraphics[width=0.8\textwidth]{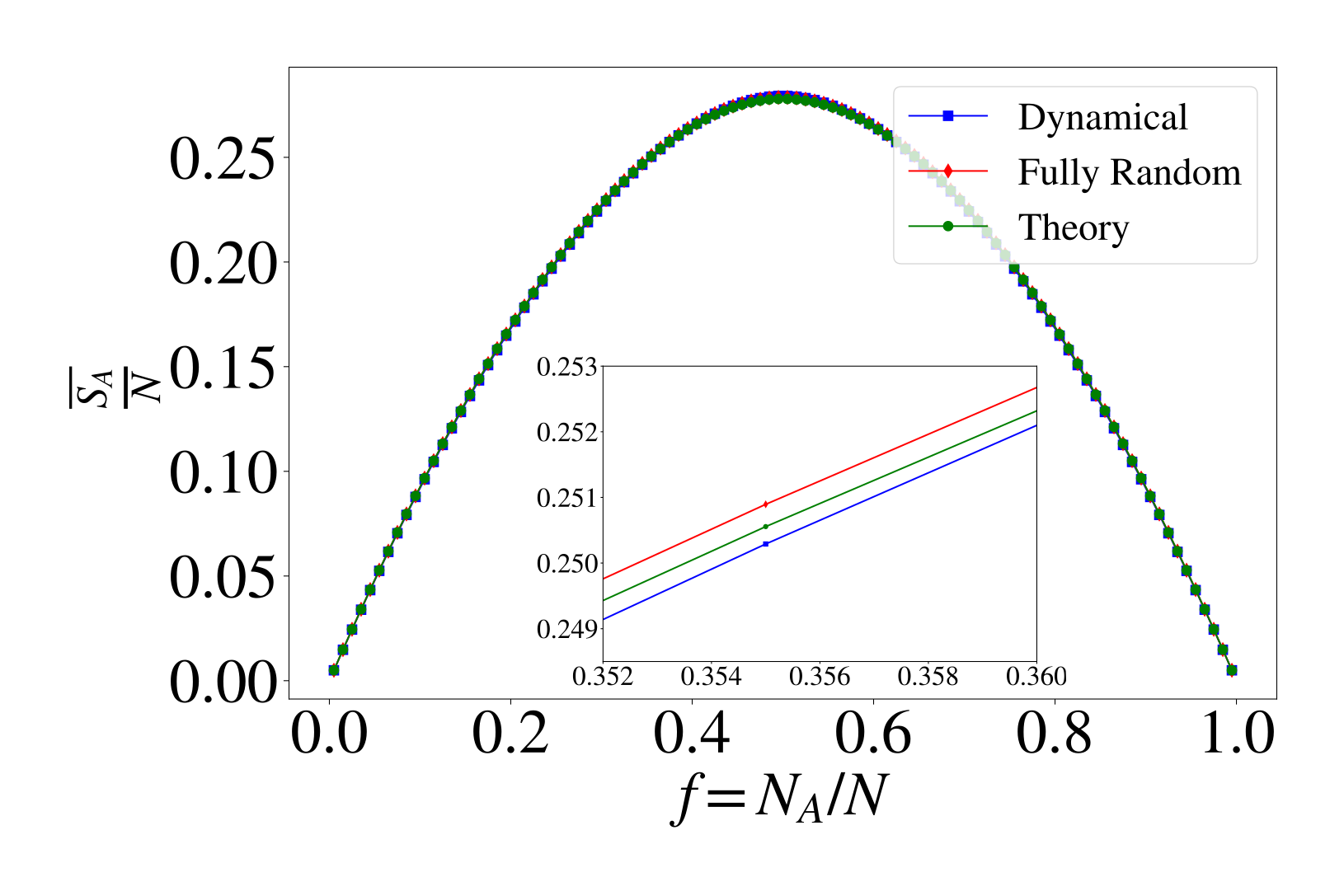}

\caption{The dynamical Page curve for Hamiltonian (\ref{eq:period2_range2_hamiltonian})
(blue curve) and its comparison with the one for the RFG ensemble (red
curve) and the theoretical result (green curve). Here $N=200$. The
theoretical result is truncated up to order $\mathcal{O}(f^{5})$,
the same as in the main text. }

\label{Taylor_completed_range3}
\end{center}
\end{figure}

In Fig.~\ref{Taylor_completed_range3},
we plotted the dynamical Page curve for the Hamiltonian 
\begin{equation}
H=\sum_{i}a_{i}^{\dagger}a_{i+1}+0.3\sum_{i:\mathrm{even}}a_{i}^{\dagger}a_{i+3}-0.3\sum_{i:\mathrm{odd}}a_{i}^{\dagger}a_{i+3}+\mathrm{H.C.}.\label{eq:period2_range2_hamiltonian}
\end{equation}
This dynamical Page curve is nearly the same as the one of minimal
model in the main text.

\subsection{Generalization to the 
atypical Page curves}

In this subsection, we 
further consider the case beyond the condition
in 
Theorem 2 in the main text, namely $\Tr(\rho Q_{k}^{\dagger}Q_{k})\neq\frac{1}{2}$
(we recall that $Q_{k+\pi}=P_{k}$). We denote $n_{k}=\Tr(\rho Q_{k}^{\dagger}Q_{k})$
and $\eta_{k}=\sqrt{n_{k}(1-n_{k})}$. Following the half filling condition,
we have 
\[
n_{k+\pi}=1-n_{k},\;\;\;\;\eta_{k+\pi}=\eta_{k}
\]
and still $a_{m}^{\dagger}=\frac{1}{\sqrt{N}}\sum_{k=0}^{2\pi}Z_{k}^{m}Q_{k}^{\dagger}e^{ikm}$
with $Z^m_{k}$ defined in Eq.~(\ref{eq:Zkm}). 
The covariance matrix
can be calculated as
\begin{align*}
[C_A]_{ml} & =\frac{1}{N}\sum_{k_{1,2}}Z_{k_{1}}^{m}Z_{k_{2}}^{l*}e^{ik_{1}m}e^{-ik_{2}l}\Tr(\rho Q_{k_{1}}^{\dagger}Q_{k_{2}})\\
 & =\frac{1}{N}\sum_{k}Z_{k}^{m}Z_{k}^{l*}e^{ik(m-l)}n_{k}+\frac{1}{N}\sum_{k}e^{i\theta_{k}}Z_{k}^{m}Z_{k+\pi}^{l*}e^{ik(m-l)}e^{i\pi l}\eta_{k}.
\end{align*}
Since $n_{k}\neq\frac{1}{2}$, in general there is no simple expression
for $X_{A}$. 

Using the same techniques as in the previous subsection, we can still
establish the Feynman rules for this case. However, we have to distinguish
two kinds of legs, one like 
$Z_{k_{j}}^{m_{j}}Z_{k_{j}}^{m_{j+1}*}e^{ik_{j}(m_{j}-m_{j+1})}n_{k_{j}}$
and the other like 
$e^{i\theta_{k_{j}}}Z_{k_{j}}^{m_{j}}Z_{k_{j}+\pi}^{m_{j+1}*}e^{ik_{j}(m_{j}-m_{j+1})}e^{i\pi m_{j+1}}\eta_{k_{j}}$.
There is no dynamical phase $e^{i\theta_{k}}$ in the former, namely no delta functions associated with contraction. Also, there is no extra $e^{i\pi m_{j+1}}$ phase term in the former.
Due to the difference between these two kinds of legs, the rule is more complicated than the previous case. 

As an example, we can calculate the first three non-trivial terms to obtain:
\begin{align*}
\overline{\Tr C_{A}^{2}} & =\frac{N_{A}}{N}\sum_{k}n_{k}^{2}+\frac{N_{A}^{2}}{N^{2}}\sum_{k}\eta_{k}^{2},
\end{align*}
\[
\overline{\Tr C_{A}^{3}}=\frac{N_{A}}{N}\sum_{k}n_{k}^{3}+\frac{3N_{A}^{2}}{N^{2}}\sum_{k}n_{k}\eta_{k}^{2},
\]
\[
\overline{\Tr C_{A}^{4}}=\frac{N_{A}}{N}\sum_{k}n_{k}^{4}+4\frac{N_{A}^{2}}{N^{2}}\sum_{k}n_{k}^{2}\eta_{k}^{2}+2\frac{N_{A}^{2}}{N^{2}}\sum_{k}\eta_{k}^{4}+\frac{2N_{A}^{3}}{N^{3}}\sum_{k}\eta_{k}^{4}-\frac{N_{A}^{4}}{N^{4}}\sum_{k}\eta_{k}^{4}.
\]
Therefore,  Up to $\overline{\Tr X_{A}^{4}}$, \footnote{Noting that in higher $n-$expansion of $\overline{\Tr X_{A}^{2n}}$, there
will also be contributions to the entropy density $\frac{\overline{S_A}}{N}$ at the order of $\mathcal{O}(f)$, like the term $\frac{N_{A}}{N}\frac{\sum_k n_{k}^{2n}}{N}$. Nonetheless, the convergence of expansion is guaranteed by $X_A^{2n}\leq X_A^{2n-2}$}
\[\overline{S_{A}}\simeq\frac{N_{A}(\ln2+\frac{3}{4})-\frac{N_{A}}{N}\sum_{k}(4n_{k}^{2}-\frac{8}{3}n_{k}^{3}+\frac{4}{3}n_{k}^{4})-\frac{N_{A}^{2}}{N^{2}}\sum_{k}(4\eta_{k}^{2}-8n_{k}\eta_{k}^{2}+\frac{16}{3}n_{k}^{2}\eta_{k}^{2}+\frac{8}{3}\eta_{k}^{4})-\frac{N_{A}^{3}}{N^{3}}\frac{8}{3}\sum_{k}\eta_{k}^{4}+\frac{N_{A}^4}{N^4}\frac{4}{3}\sum_{k}\eta_k^4}{\ln2}.
\]

\section{Calculation of entanglement Entropy in the Quasi-Particle Picture}

In the quasi-particle picture, a nonequilibrium 
initial state is a source for generating quasi-particles
with opposite momenta, which travel ballistically through the system.
Here the main assumption is 
 those quasi-particle pairs generated 
at different locations and times are incoherent. Therefore, the entanglement entropy
of subsystem $A$ is proportional to the number of pairs shared between
$A$ and its complement. Without loss of generality, we can assume
the subsystem $A$ is located in $[0,N_{A})$, $N_{A}\leq\frac{N}{2}$.
For a certain type of pairs with velocity $\pm v(k)$ ($v(k)>0$),
if the right-end of the pair is at position $0\leq x<N_{A}$, i.e.
within subsystem $A$, only when $x$ satisfies
\[
N_{A}-N\leq x-2v(k)t<0
\]
can this pair contribute to the entanglement entropy of $A$. Here the periodic
boundary condition is taken into account and $2v(k)t$ should be understood
as modulo $N$. The solutions of this inequality is a continuous range
$x\in[x_{\mathrm{min}},x_{\mathrm{max}})$, where $x_{\mathrm{min}}=\max\{0,2v(k)t+N_{A}-N\}$ and 
$x_{\mathrm{max}}=\min\{N_{A},2v(k)t\}$. Accordingly, we obtain
\[
\Delta x=x_{\mathrm{max}}-x_{\mathrm{min}}=\begin{cases}
2v(k)t, & 2v(k)t\leq N_{A};\\
N_{A}, & N_{A}<2v(k)t<N-N_{A};\\
N-2v(k)t, & 2v(k)t\geq N-N_{A}.
\end{cases}
\]
A similar argument holds if the left-end is in subsystem $A$. We
assume that after a sufficiently long 
time, the quasi-particle pairs will distribute
uniformly among the system. Hence, 
the contribution of 
quasi-particle pairs with momentum $k$ to the entanglement entropy upon the long-time average is given by 
\[
\frac{S_{A}(k)}{N^{2}}\int_{0}^{N}d(2v(k)t)\Delta x=S_{A}(k)\left[\frac{N_{A}}{N}-\left(\frac{N_{A}}{N}\right)^{2}\right],
\]
where the coefficient $S_A(k)$ is to be determined. Summing over all types of pairs, we obtain 
\[
S^{\rm qp}_{A}=\left(\frac{N_{A}}{N}-\frac{N_{A}^{2}}{N^{2}}\right)\sum_{k}S_{A}(k).
\]
If $N_{A}\to0$, the limit $S_{A}^\mathrm{qp}\to N_{A}\sum_{k}\frac{H(n_{k})}{N}$
should hold \citep{Alba2018}. Therefore, $S_{A}(k)=H(n_{k})$. If
$n_{k}=\frac{1}{2}$ for all $k$'s, the entanglement entropy for subsystem
$A$ is
\[
S^{\rm qp}_{A}=N_{A}-\frac{N_{A}^{2}}{N},
\]
which deviates considerably from 
the dynamical Page curve 
discussed in the main text.

\end{document}